\documentclass[preprint2]{aastex}

\newcommand{\feh}{\ensuremath{[\mbox{Fe}/\mbox{H}]}}
\newcommand{\rphk}{\ensuremath{R'_{\mbox{\scriptsize HK}}}}
\newcommand{\fc}{\ensuremath{F_{\mbox{\scriptsize\ion{Ca}{2}}}}}
\newcommand{\dfc}{\ensuremath{\Delta\fc}}
\newcommand{\caii}{\ion{Ca}{2} H and K}
\newcommand{\mv}{\ensuremath{M_{\mbox{\scriptsize V}}}}
\newcommand{\ms}{\ensuremath{M_{\mbox{\scriptsize V,MS}}}}
\newcommand{\dms}{\ensuremath{\Delta M_{\mbox{\scriptsize V}}}}
\newcommand{\fcmin}{\ensuremath{F_{\mbox{\scriptsize\ion{Ca}{2},min}}}}
\newcommand{\teff}{\ensuremath{T_{\mbox{\scriptsize eff}}}}
\newcommand{\sv}{\ensuremath{\sigma_{rv}}}
\newcommand{\smeas}{\ensuremath{\sigma_{\mbox{\scriptsize meas}}}}
\newcommand{\sj}{\ensuremath{\sv^\prime}}
\newcommand{\lrphk}{\ensuremath{\log{\rphk}}}
\newcommand{\myemail}{jtwright@astro.berkeley.edu}
\newcommand{\persec}{\ensuremath{\mbox{s}^{-1}}}

\slugcomment{Draft of an upcoming paper for PASP}

\shorttitle{Velocity Jitter in Planet Search Stars}
\shortauthors{Wright}

\begin{document}
\title{Radial Velocity Jitter in Stars from the California and Carnegie
  Planet Search at Keck Observatory\altaffilmark{1}}
\altaffiltext{1}{Based on observations obtained
at the W. M. Keck Observatory, which is operated jointly by the
University of California and the California Institute of Technology.
The Keck Observatory was made possible by the generous financial support of the W.M. Keck Foundation.}
\author{J. T. Wright}

\affil{Department of Astronomy, 601 Campbell Hall, University of California, Berkeley, CA 94720-3411}
\email{\myemail}

\begin{abstract}
I present an empirical model for predicting a star's radial velocity
jitter from its \bv\ color, activity level, and absolute magnitude.
This model is based on observations of $\sim 450$ well-observed stars
from Keck Observatory for the California and Carnegie Planet Search
Program.  The model includes noise from both astrophysical sources and
systematic errors, and describes jitter as generally increasing with
a star's activity and height above the main sequence. 

\end{abstract}


\section{Introduction}

The California and Carnegie Planet Search has been observing $\sim 1200$
stars, some since 1997, at Keck Observatory in an ongoing program to
detect radial velocity variations caused by planetary companions.
One source of error in the measured velocities is ``jitter'' due, in
part, to flows and inhomogeneities on the   
stellar surface that can produce variations in the measured radial velocity
of a star, and can even mimic the signature of planetary
companions (\citet{Henry2002,Queloz2001,Santos2003}).
\citet{Marcy2005} describe how the California and Carnegie Planet
Search distinguishes planetary signals from jitter.

The best-fit orbital parameters of a Keplerian velocity model depend
on the relative errors of points in an observed 
radial velocity time series.  Thus, one requires a quantitative model
of all sources of noise, including jitter, in order to properly
calculate the orbits of a potential planetary companion.

\citet{Saar1997} studied jitter in F, G, and K dwarfs, modeling the
effects of spots and convection on radial velocity measurements as a
function of \teff\ and activity.  They concluded that 
radial velocity jitter should increase with \teff, activity, and $v
\sin{i}$, and found this to be the case empirically in a small sample of
active stars. \citet{Paulson2004} found that among four Hyades stars with
significant radial velocity variation, three showed correlated
photometric variation, consistent with the effects of rotationally
modulated spots and plage.  \citet{SaarButler1998} used a sample from the Lick
Planet Search results to produce a formula for predicting jitter in
stars given $P_{\mbox{rot}}$, $v\sin{i}$, $v_{\mbox{mac}}$, and
photometric variability.

The purpose of this work is not to explore the astrophysical sources of
jitter or to refine the formula of \citet{SaarButler1998} which
predicts jitter from photometric variability and $v\sin{i}$ (a
quantity that is difficult to measure in stars as old as typical
Planet Search stars),  but to provide an empirical estimate of jitter
using stellar parameters available without an independent photometric
program or detailed spectroscopic analysis: \mv, \bv, and $S$.  The
strategy is to divide the Planet Search sample into groups of
stars with similar values for these parameters.  The model then simply
predicts that a given star will exhibit levels of jitter similar to
those stars with similar properties, i.e., in the same group.

\section{Data}
\subsection{The Sample\label{sample}}
To measure the radial velocity jitter this work employs a sample of
stars that have useful radial-velocity time
series and no detected companions. From the $\sim 1200$ stars in the
Keck Planet Search program, 
this sample comprises 531 stars with at least nine radial velocity
measurements, with activity levels reported in \citet{Wright2004}, and
{\it Hipparcos} colors and parallaxes \citep{Hipparcos}.   

Rather than attempt to subtract the velocity signatures of known companions
from the sample, some of which may be rather poorly determined, the
sample here excludes those stars harboring published or announced companions 
from the sample altogether. Furthermore, this sample excludes stars that
show clear, unambiguous evidence for unannounced companions (stellar and
planetary).  These cases involve very long period companions of
unknown mass where the radial velocities show curvature, or Keplerian
amplitudes of more than 1 km\persec.  Finally, the sample excludes several
``borderline'' stars whose radial-velocity time series have best-fit
Keplerians with false alarm probabilities \citep{Marcy2005} of less
than 0.1.  

The final sample comprises 448 stars.

\subsection{The Evolution Metric\label{dms}}

\begin{figure}
\plotone{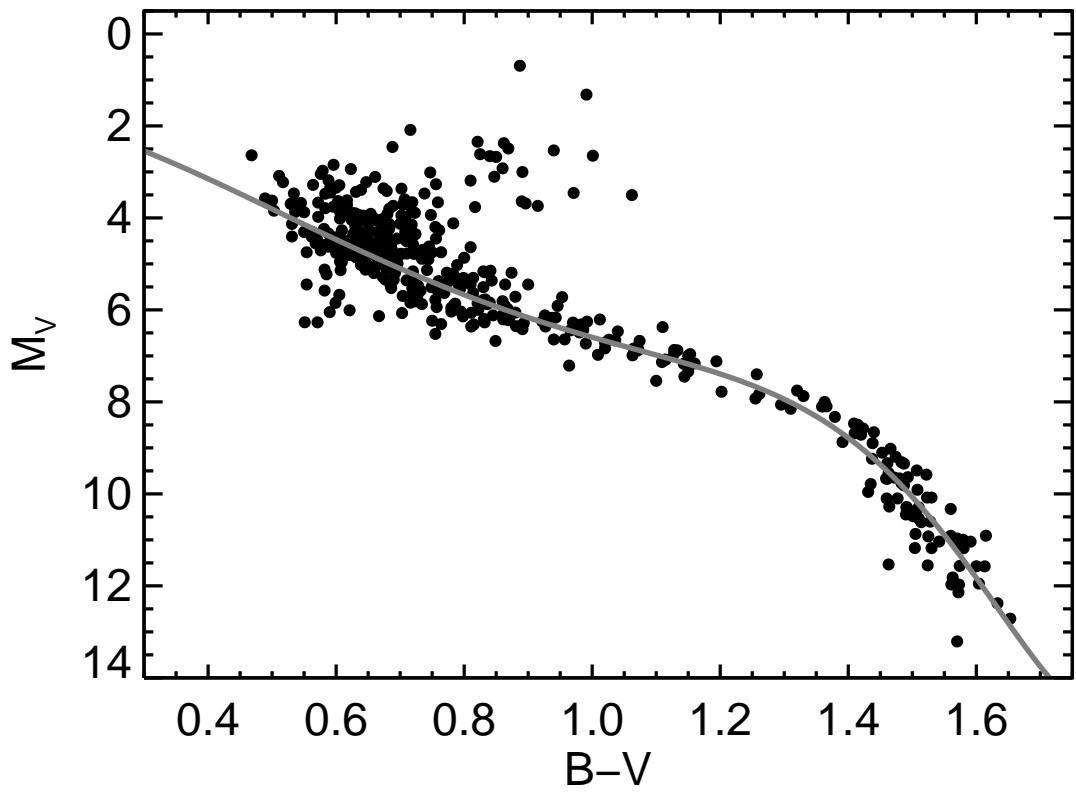}
\caption{Color-magnitude diagram for the stars used in this work.  The
  solid line is the fit to the main sequence used to determine \dms,
  the height above the main sequence.\label{msfig}}
\end{figure}

Following \citet{Wright2004b}\notetoeditor{this reference has been
  corrected in an erratum to be published in the March 2004 AJ.  Since
the material cited here is affected by this erratum, it may be
appropriate to include the erratum in the reference, either here or in
the references section.  Is there a standard way of doing this?}, this work employs the evolution metric

\begin{equation} 
\dms = \ms(\bv) - \mv
\end{equation}  

as a quantitative proxy for evolution.  Here,  \mv- and
\bv-values are from the {\it Hipparcos} catalog, and

\begin{equation}
\label{ms}
\ms(\bv) = \sum a_i (\bv)^i
\end{equation}

represents the {\it Hipparcos} average main sequence where $a=\left\{\right.$1.11255,
5.79062, -16.76829,  76.47777,-140.08488, 127.38044, -49.71805,
-8.24265,  14.07945,  -3.43155$\left.\right\}$.  Fig.~\ref{msfig}
shows \ms\ plotted on a color-magnitude diagram with the stars
from the sample described in \ref{sample}.   Evolved stars, lying
above the main sequence, have positive values of \dms, while subdwarfs
have negative values.   

While a more appropriate metric might employ measured abundances to lift the
degeneracy between the effects of metallicity and evolution on \dms,
the goal here is to provide a description of radial velocity jitter
without the benefit such measurements.

\subsection{The Activity Metric}

Active stars, especially active F dwarfs, exhibit higher levels of
jitter than their inactive counterparts.  To quantify this
relationship, this work employs the activity metric $S$, as measured in
\citet{Wright2004} for the Planet Search stars, which represents the
amount of emission in the \caii\ line core normalized by nearby
continuum regions.  \citet{Wright2004}, following \citet{Noyes1984},
transformed $S$ to \rphk\, which represents the fraction of
a star's total luminosity emitted by the chromosphere in the \caii\
lines.  Example spectra containing \caii\ lines can be found in most
of the planet discovery papers by the California and Carnegie Planet Search,
e.g. \citet{Butler2003}. 

The quantity \rphk, however, is not well calibrated for stars with
$\bv < 0.4$ or $\bv > 0.9$.  To overcome this, this work employs an alternative
metric of stellar activity, \fc, defined by \citet{Rutten1984} (who
used the expression 
``$F_{\mbox{\scriptsize{H}}} + F_{\mbox{\scriptsize{K}}}$'' in that
work) and described again in \citet{Rutten1986}.  This metric
differs from \rphk\ primarily in that it does not correct for a
photospheric component to the \caii\ lines and in that it is a flux density,
rather than a fraction, since it is constructed using an additional
factor of $\teff^4$.  \citet{Rutten1987b} argued that \fc\ is a more
appropriate metric for comparing activity levels among stars since it
is related to 
soft X-ray and \ion{Mg}{2} h and k fluxes by power laws with no color
or luminosity dependence.  \citet{Hall1995} determined the units of
\fc\ to be $7.97 \times 10^5 \mbox{ erg} \mbox{ cm}^{-2}
\mbox{s}^{-1}$.  \citet{Rutten1984} prescribes 

\begin{equation}
\fc=S C_{\mbox{\scriptsize cf}} \teff^4 10^{-14}
\end{equation}

to transform the $S$-values of the stars in the sample here to \fc\ where 

\begin{equation}
\log{C_{\mbox{\scriptsize cf}}} =  
	           0.25  (\bv)^3
		  -1.33  (\bv)^2
                  +0.43  (\bv)
	          +0.24 
\end{equation}
for main sequence stars with $0.3 \leq \bv \leq 1.6$.  This work takes
\teff\ for sample stars from \citet{Fischer2005} or, for the few stars
here without such measurements, from the \teff-\bv relation of
\citet{Flower1996}. 

\begin{figure}
\plotone{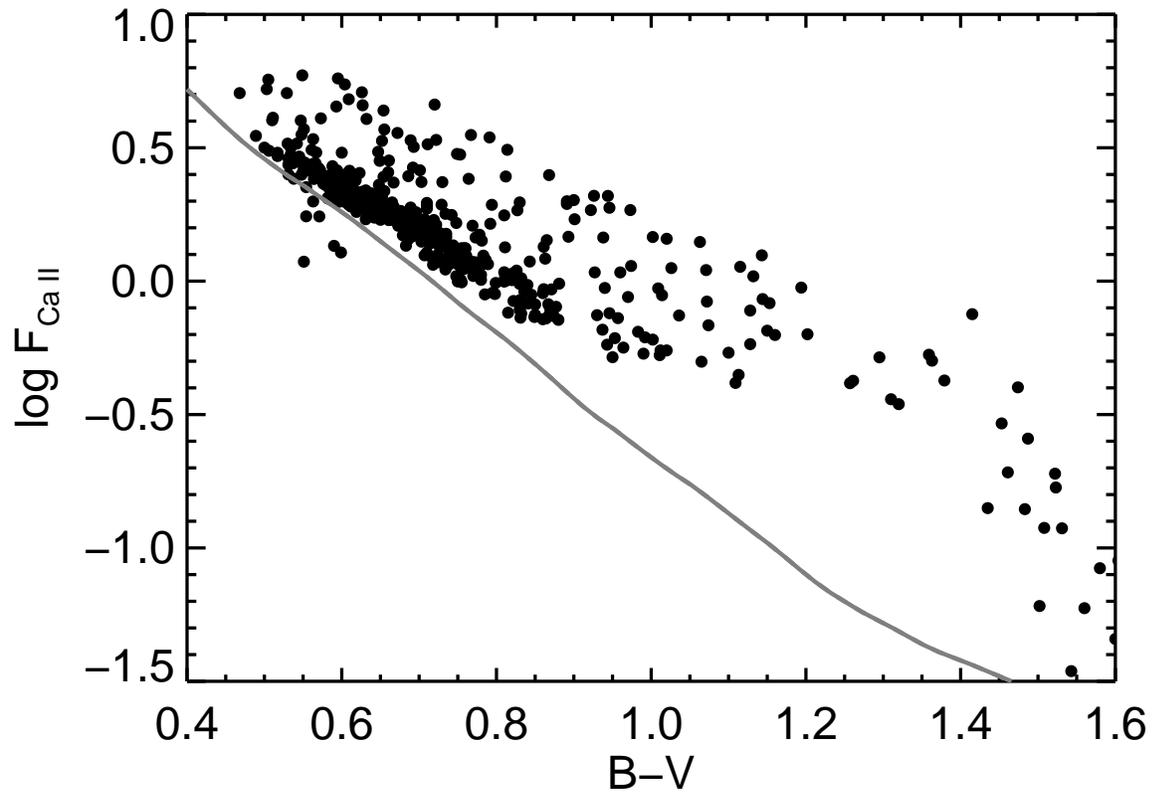}
\caption{Activity, \fc, as a function of color for all of the stars used in
  this work.  The solid line represents the minimum activity for
  dwarfs, \fcmin.\label{fbv}}
\end{figure}

Fig.~\ref{fbv} shows \fc\ as a function of color for the sample
stars here.  \citet{Rutten1987} tabulated the empirical minimum activity
level \fcmin\ observed for stars as a function of \bv.  
Table~\ref{lowerrutten} and the solid line in Fig.~\ref{fbv}
give this tabulation, converted to the units of this work.  This
 minimum level  ostensibly represents radiative flux not related to
 magnetic activity.  A comparison of Fig.~\ref{fbv} with Fig.~6a of
 \citet{Rutten1984} shows good agreement (note that the curves in
 Rutten's figure have been superseded by the curves referenced above).

This work employs the activity metric \dfc, constructed as:

\begin{equation}
\dfc = \fc - \fcmin
\end{equation}

For some  metal-poor, blue stars, \dfc\ is negative.
Apparently, these stars' very low metallicities ($\feh < -1$)
require either a different calibration of \fc\ or a different
expression for \fcmin\ (or both).  This is not a problem in this
analysis, because, as described below, \dfc\ is not needed to predict
the jitter for stars as inactive as these. 

\begin{table}
\begin{center}
\caption{The lower boundary of the surface flux \fc\, as a function of
  \bv, adapted from \citet{Rutten1987} to the units of this
  work.\label{lowerrutten}} 
\begin{tabular}{cc}
\tableline\tableline
\bv & \fcmin \\
\tableline
0.4 & 5.24 \\
0.45 & 3.80 \\
0.5 & 2.88 \\
0.55 & 2.29 \\
0.6 & 1.82 \\
0.65 & 1.41 \\
0.7 & 1.09 \\
0.75 & 0.831 \\
0.8 & 0.645 \\
0.85 & 0.489 \\
0.9 & 0.363 \\
0.95 & 0.281 \\
1.0 & 0.218 \\
1.05 & 0.174 \\
1.1 & 0.135 \\
1.15 & 0.105 \\
1.2 & 0.079 \\
1.25 & 0.063 \\
1.3 & 0.052 \\
1.35 & 0.044 \\
1.4 & 0.038 \\
1.5 & 0.029 \\
1.6 & 0.025 \\
\tableline
\end{tabular}
\end{center}
\end{table}

Henceforth, this work uses the term ``active'' to refer to stars
with $\fc > 0.6$.  This limit corresponds to $\lrphk \sim -4.95$ at $\bv = 
0.5$ and $\lrphk \sim -4.8$ at $\bv = 0.9$.  

\subsection{Radial Velocity Measurements}
\subsubsection{The Jitter Metric}

Many sample stars exhibit linear trends in their radial
velocity time series due to long-period planetary or stellar
companions.  To remove such effects, the standard deviation of
the velocities, \sv\ is defined about a linear trend, $v_r=a+bt$, for all of
the time series:

\begin{equation}
\sv^2=\sum_{i=0}^N \frac{(v_{r,i}-a-b t_i)^2}{N-2} 
\end{equation}

The quantity \sv\ includes random measurement uncertainties such as
those arising from photon statistics, estimated as follows.  Each
observation of a star consists of a spectrum divided 
into ~400 chunks, each of which is analyzed independently the Planet
Search Doppler analysis pipeline to determine 
the value of $\Delta v$, the differential radial velocity of the star
with respect to a template observation.  The distribution of the
$\Delta v$ values of these 400 chunks provides an estimate of the
measurement error for a given observation.  Because the Planet Search
employs an exposure meter at Keck, most observations of a given star have the
same signal-to-noise (S/N) ratio, so the measurement errors for all
observations of a given star are usually very similar.  Thus, a single
number \smeas\ provides an estimate of the contribution of measurement
error to \sv\ for observations of a given star. 

Following \citet{SaarButler1998}, this work employs the metric 
$\sj\ = (\sv^2 - \smeas^2)^{1/2}$ to quantify radial velocity
variations in excess of \smeas.
The exposure meter at Keck ensures that those velocity errors limited
by S/N are roughly constant across the sample,
contributing 2-4 m\persec to the error budget.  The exceptions to this
rule are faint M dwarfs, which receive a maximum of $\sim 10$ minutes
per exposure and therefore have photon-limited errors of 3-8 m\persec, and
active F dwarfs that have high astrophysical jitter 
and receive short exposure times to keep errors photon-limited.  For
these stars, the observed rms velocities are nearly equal to
\smeas, indicating that photon statistics dominate the error budget.
In cases in which where radial velocity variations in excess of \smeas\ remain
undetected, this work simply assigns an upper limit of $\smeas/2$:

\begin{equation}\label{jitterform}
\sj           \left\{ \begin{array}{ll}
	           = \sqrt{\sv^2-\smeas^2} & \sv^2 > \frac{5}{4}
	           \smeas^2 \\
               \\
		   < \smeas/2              & \sv^2 \leq \frac{5}{4} \smeas^2 \\
                      \end{array}   \right.
\end{equation}

\subsubsection{The Components of \sj\label{errors}}

In addition to astrophysical jitter, \sj\ also measures undetected
planetary companions and systematic errors as discussed below.

Many stars in this sample undoubtedly harbor still-undetected,
low-amplitude planets.  Only after many planetary orbits with good
phase coverage can one use 
the periodic nature of the signals and Keplerian fits to distinguish
these signals from jitter \citep{Marcy2005}.  Refining these fits to
determine the precise orbital elements of a planetary companion
requires still more observations, and the statistics of the residuals
to these fits is complex.  It is for the sake of simplicity and uniformity
in the sample that this work employs only linear fits to the velocity
time series and not Keplerians. 

Systematic errors have many sources, including imperfect deconvolution
of template spectra (especially for M dwarfs) and poor charge transfer
in the HIRES Tek CCD.  The Planet Search estimates these errors to be on
the order of 3 m\persec for M dwarfs and $< 2$ m\persec for other stars
(G. W. Marcy 2005, private communication).  Some
component of these systematic errors will be accounted for which the
subtraction of \smeas\ from \sv, but systematic errors, by their
nature, are not well characterized by Gaussian noise, so some component
of these errors probably remain in \sj.

\section{Analysis}

The purpose of this paper is to quantify the known dependencies of
radial velocity jitter on stellar parameters, namely that active, blue, and
evolved stars exhibit more jitter than old, main sequence, G
dwarfs.  This section, therefore, discusses trends of jitter with
respect to \dfc, \dms, and \bv.

\subsection{Activity\label{act}}
\begin{figure}
\plotone{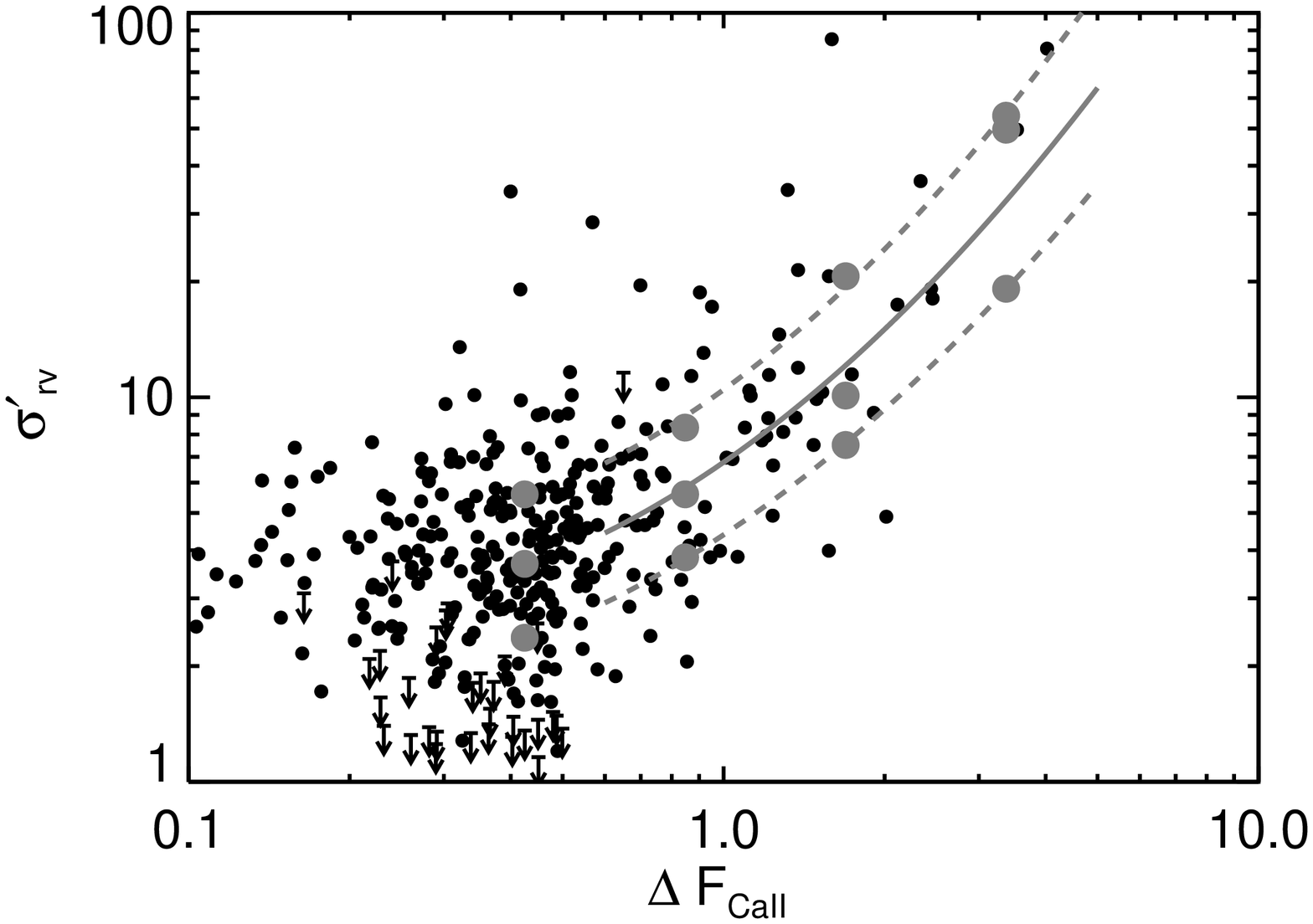}
\caption{Activity, \dfc, vs. the observed jitter, \sj\ for stars with
  $\dms < 1$.  Typical errors
  on data points are around 1.5 m\persec.  Upper limits
  are represented by arrows.  Large gray disks represent the 20th
  percentile, median, and 80th percentile in each of 4 equally spaced
  bins in $\log{\dfc}$ from $0.3 \leq \dfc \leq 3$.  The dashed lines
  represent the parabolic fits to these points in the range $0.6 \leq
  \dfc \leq 3$, where the model is applicable.\label{actjit}}
\end{figure}

Fig.~\ref{actjit} shows \sj, as a function of \fc\ for stars with $\dms < 1$.  There is a clear increase in \sj\ with \fc, such that
the most active stars consistently show more than 4 m\persec of
jitter.  A few stars with $\dfc < 1$ appear to show more than $\sim
20$ m\persec of jitter;  these stars may harbor planets whose orbital
periods have not yet become apparent from the radial velocity time
series.  However, two of these stars may be very young ($< 1$ Gyr)
according to \citet{Fischer2005}, so one cannot discount the
possibility that these \sj values are due purely to astrophysical jitter.

This work employs the following method to fit a polynomial to these
data, which contain a few high outliers and points for which only an
upper limit can be set.  First, I divide the distribution into four evenly
spaced bins in $\log{\dfc}$.  Within each bin I then find the median value,
20th percentile, and 80th percentile of $\sj$ (represented
by large gray filled circles in Fig.~\ref{actjit}), yielding three sets of
points to which I fit second-order polynomials. In
Figure 3, dashed lines represent the fits to the 20th and 80th percentile
points, and the solid line represents the avereage of these fits,
representing the model of the median.
  The Appendix describes the definition of a median and
a percentile in a data set containing upper limits.  Note that the choice of a
second-order polynomial here is not physically or
statistically motivated, but is merely chosen to be a simple, accurate
description 
of the distribution with a smooth transition to the low-activity subsample.

There is very little change in the median jitter for stars with $\dfc
< 0.6$, so these fits are only applicable for stars with $\dfc > 0.6$.  Of 
the 72 such stars in the sample, 53\% lie below the solid line in Fig.~\ref{actjit},
80\% lie below the 80th percentile fit, and 20\% lie below the 20th
percentile fit.  Table~\ref{jitters} contains coefficients for these
polynomials. 

The result here, that jitter increases with activity is consistent with
the result of \citet{Saar2003}, who found $\sj \propto (\rphk)^b$, with
$b=1.1$ for G and K dwarfs and $b=1.7$ for F dwarfs.  Although here
the fit is parabolic and is applied to a different metric, one can
compare the results by fitting a line to \sj\ versus \lrphk\ for the F
stars and the G and K stars separately, using the ASURV statistical
analysis package rev 1.2 \citet{asurv}, which implements the methods
presented in \citet{Isobe1986} to handle the upper limits.  The
best-fit values for found $b$ by applying the Buckley-James method and
the parametric 
EM algorithm are $b=1.6$ for F dwarfs ($\bv < 0.63$,$\dms < 1$) and
$b=0.80$ for G and K dwarfs ($0.63 < \bv <1.4$, $\dms < 1$).  While
these results differ somewhat from the results of Saar et al., the
strong color dependence remains. 

\begin{figure}
\plotone{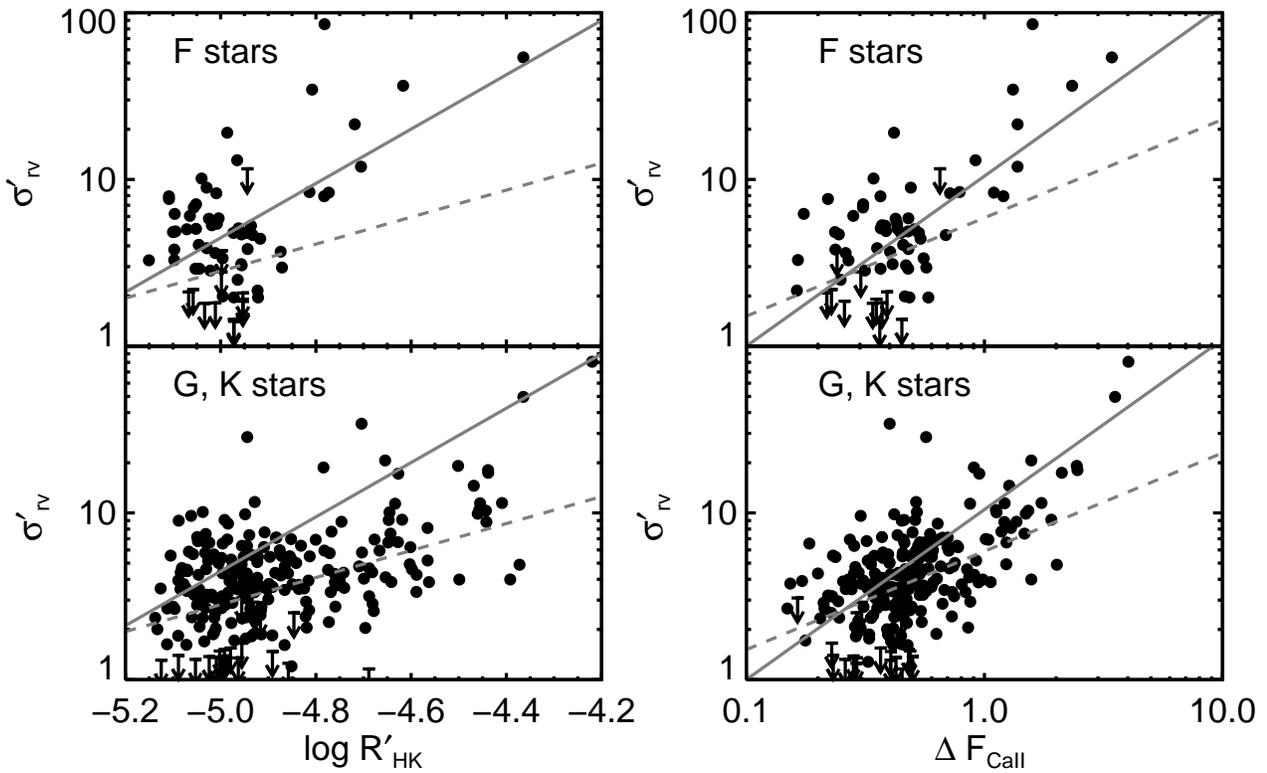}
\caption{Jitter as a function of activity using two different activity
  metrics.  Solid lines represent fits
using only F stars, dashed lines those using only the G and K stars.
  F stars show a 
  steeper activity dependence than G and K stars with \lrphk\ as a
  metric (left panel), but the difference is smaller when \fc\ is
  the metric (right panel).  These fits are provided for comparison
  only, and are not used in the jitter model presented
  here.\label{saarplot}}  
\end{figure}

However, using the activity metric \dfc\ instead, however, with its
extra factor of 
$\teff^4$, this color dependence is somwhat reduced:  a similar fit to the
relation $\sj \propto (\log{\dfc})^c$ returns $c=0.59$ for G and K
dwarfs and $c=1.01$ for F dwarfs.  The choice of \dfc\ thus reduces the color dependence of the
jitter-activity relation, as shown in Fig.~\ref{saarplot}.

\subsection{Stellar Evolution\label{evo}}
\begin{figure}
\plotone{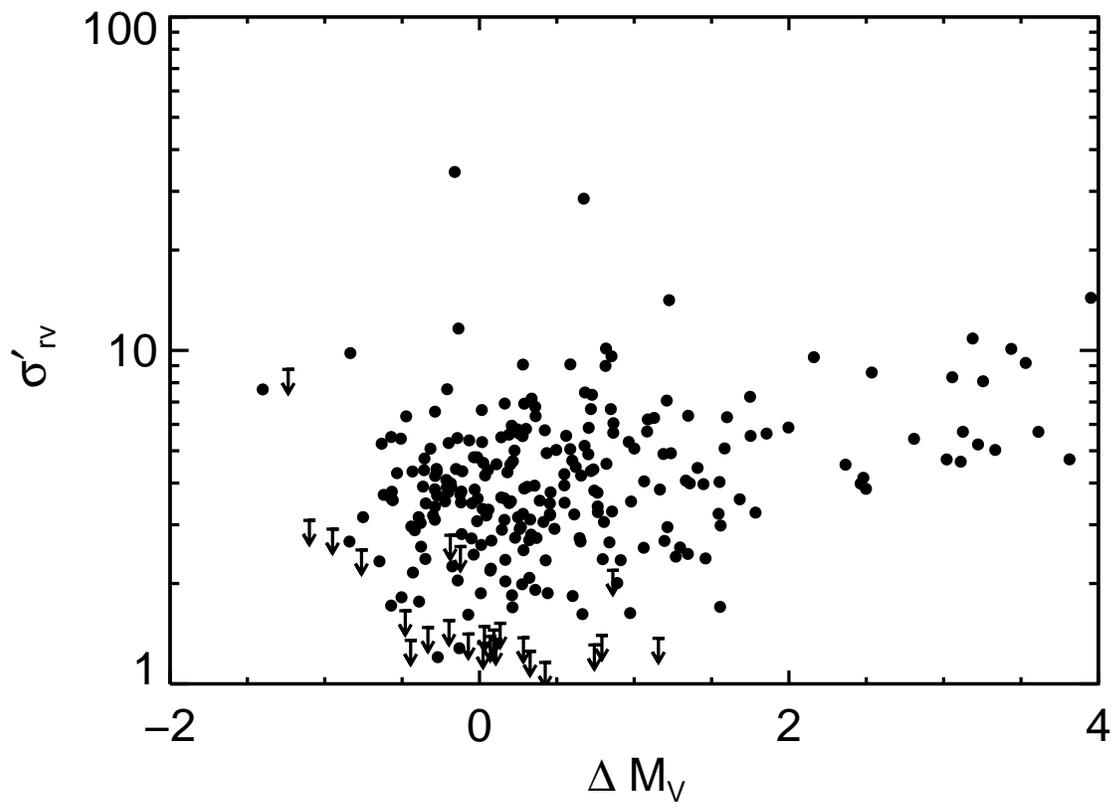}
\caption{Evolution, measured as \dms\ vs. \sj, for stars with $\fc < 0.6$.  Typical errors on data points are around 1.5 m\persec.
  Arrows represent upper limits.  The most evolved 
  stars exhibit large amounts of jitter.\label{evojit}}
\end{figure}

Fig.~\ref{evojit} shows \sj, as a function of \dms\ for stars with
$\dfc < 0.6$.  Stars with $\dms > 2$ have higher typical
\sj-values than stars on the main sequence.  The trend here is weak,
so in this case, the model consists of three broad groups, each modeled
independently: main-sequence stars ($\dms < 1$), 
slightly evolved stars ($1 < \dms < 1.75$), and subgiants ($\dms >
1.75$).

\begin{figure}
\plotone{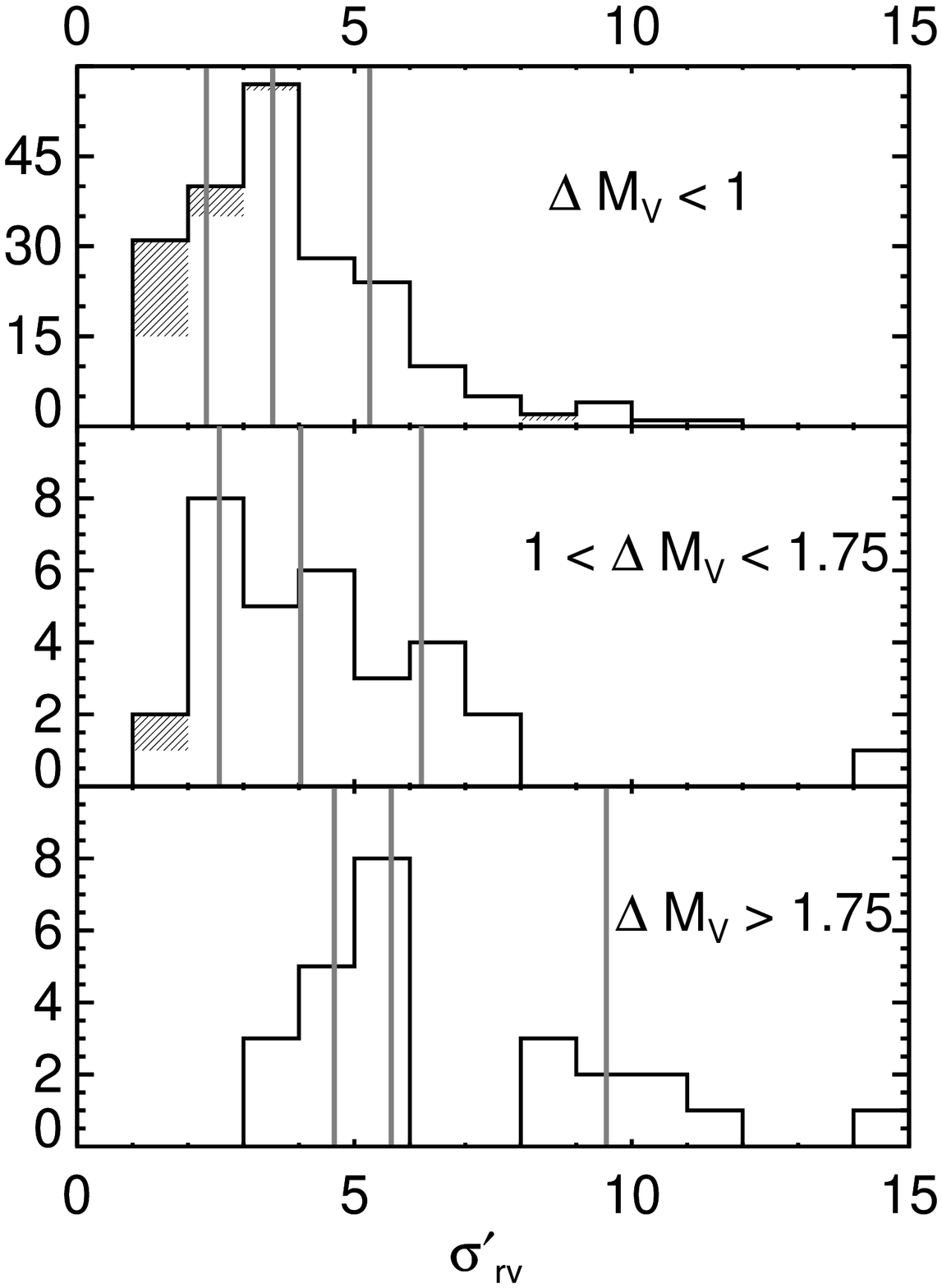}
\caption{Distribution of \sj, the observed jitter, for inactive ($\dfc
  < 0.6$) stars in bins of \dms.  The top panel includes only stars
  for which $0.6 < \bv < 1.4$, matching the majority of stars in the bottom
  two panels.  Shaded regions indicate stars for which only upper
  limits on the jitter can be set.\label{evohist}}
\end{figure}

Fig.~\ref{evohist} shows the distribution of \sj\ values for these
three bins in \dms, with shaded regions representing those stars for
with \sj\ exceeds $\smeas/2$ (that is, upper limits on \sj) (Eq.~\ref{jitterform}).  The median,
20th, and 80th percentile levels for these distributions appear as vertical
gray bars in Fig.~\ref{evohist}, and numerically in
Table~\ref{jitters}. Again, the appendix describes the
statistics of medians for these data sets.

Fig.~\ref{evohist} shows that subgiants exhibit higher \sj\ values
than their main-sequence 
counterparts, having a median \sj\ value $> 5$m\persec, significantly
higher than the non-astrophysical sources of noise discussed in
\S~\ref{errors}.  In fact, no subgiant in the sample performs better
than 3 m\persec.  These stars are therefore poor velocity standards,
due to jitter.

The distribution of slightly evolved stars is very similar to that of
main sequence stars.  This work nonetheless treats them
separately here, because their activity levels, normally a good
indicator of jitter, are generally lower and therefore not comparable
to main-sequence stars \citep{Wright2004b}.  Their numbers
are too few for further subdivisions into bins of color or activity,
so the model given here simply incorporates median \sj\ values for this
group as a whole.

\subsection{Color\label{color}}

\begin{figure}
\plotone{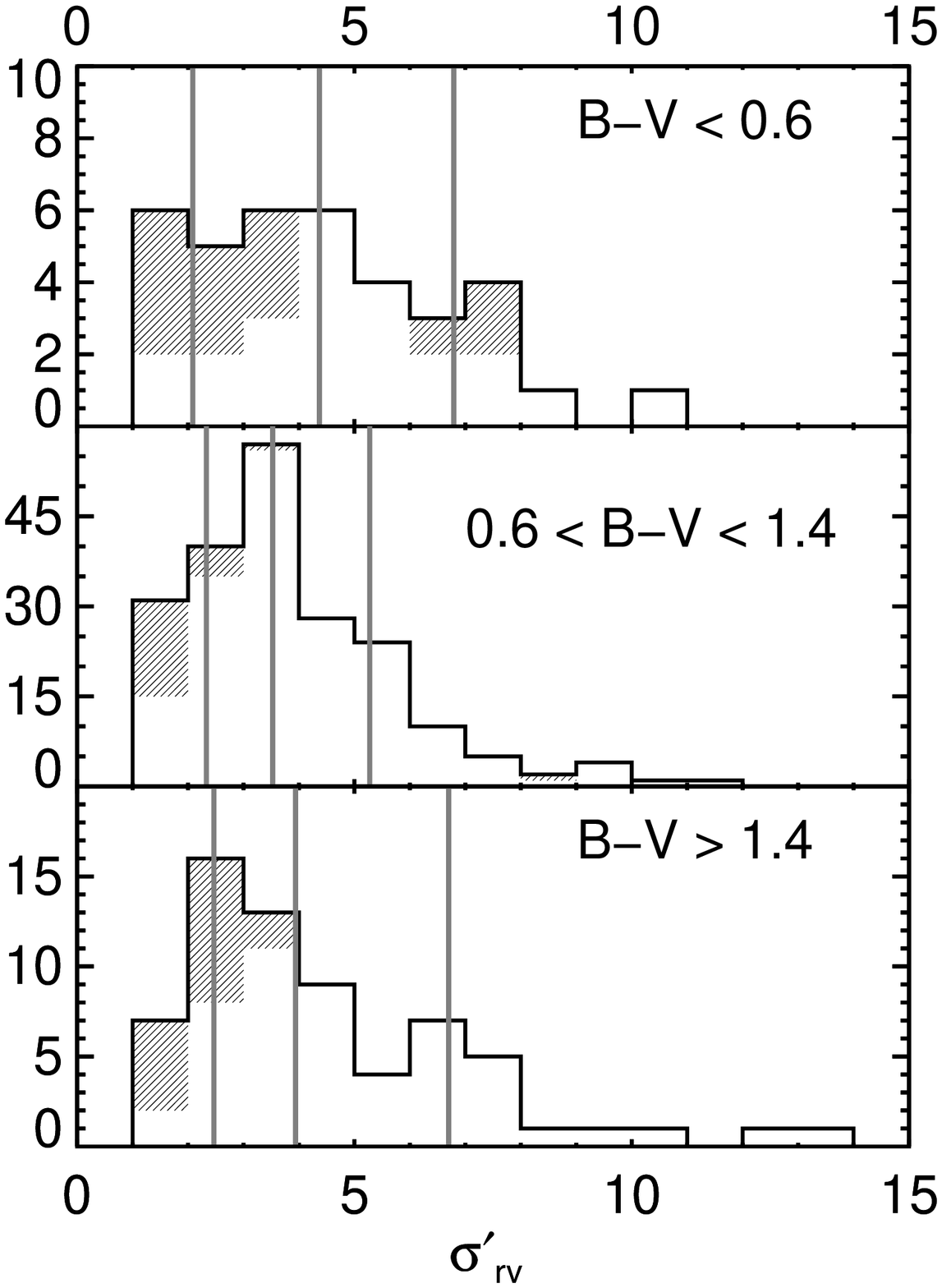}
\caption{Distribution of \sj, the observed jitter, for inactive ($\dfc
  < 0.6$), unevolved ($\dms < 1$) stars in three color bins.  The
  intermediate color stars show the lowest median jitter.  Shaded
  regions indicate stars for which only upper limits on the jitter can
  be set.\label{normhist}} 
\end{figure}

Fig.~\ref{normhist} shows the distribution of \sj\ values for stars
in three \bv\ ranges (corresponding roughly to spectral types F, G and
K, and M) that have $\dms < 1$ and $\fc < 0.6$. Here again,
vertical bars show the values of the 20th percentile, median,
and 80th percentile, and shaded regions represent upper limits.

Jitter in these unevolved inactive stars is not a well-determined, smooth
function of color, but, broadly speaking, three distinct \bv\ groups
exist.  Blue stars ($\bv < 0.6$) generally show $\sim 5$ m\persec of
jitter, while G and K dwarfs exhibit a median \sj\ value of 3 m\persec.
The large number of shaded stars in the distribution of F stars in
Fig.~\ref{normhist} reflects the fact that, as noted in
\S~\ref{sample}, Planet Search stars do not receive more exposure time
than their jitter levels warrant, to keep errors photon-limited.  Note
that this practice does not artificially inflate the estimate of 
jitter in these stars because the method of calculating medians
accounts for these upper limits (Appendix).

The \sj\ distribution of M dwarfs exhibits a tail of high
jitter values ($\sj \sim$ 7 m\persec), which most 
likely represents systematic errors in the deconvolution of their
spectra, made difficult by the lack of continuum in such cool stars.
The large shaded regions of the M dwarf distribution are due to these
stars' faintness: as noted in \S~\ref{sample}, no Planet Search star
receives more than 10 minutes of exposure at a time, so faint M dwarfs
will yield lower S/N ratios and thus have photon-limited
precision. 

G and K dwarfs yield the best radial-velocity performance, with over
30 members exhibiting rms velocities below 2 m\persec.  It is 
fortunate that the best performing stars in the sample are those that
most closely resemble the Sun.   

\begin{deluxetable}{cccccc}
\tablecaption{Jitter values for different regions in \dfc, \bv, \dms\
  space.  Values in curly braces are coefficients in a polynomial fit:
   for an entry $\{a_0,a_1,a_2\}$, $\log{\sj} = \sum_i a_i (\bv)^i$. \label{jitters}} 
\notetoeditor{The caption and label of the table appear as small caps
  in my version of LaTeX, inconsistent with Table 1.  I do not know
  why;  it may be a product of the deluxetable macro.  At any rate,
  this is unintentional.}
\tabletypesize{\scriptsize}
\tablecolumns{6}
\tablehead{
\multicolumn{3}{c}{Region} & \multicolumn{3}{c}{\sj (m\persec)} \\
\colhead{\dms} & \colhead{\dfc} & \colhead{\bv} & \colhead{20th percentile} & \colhead{median} & \colhead{80th percentile}}
\startdata
$1.75 < \dms < 4$ & $< 0.6$ & \nodata & 4.6 & 5.7 & 9.5 \\
$1 < \dms < 1.75$ & $< 0.6$ & \nodata & 2.6 & 4.0 & 6.2 \\
$< 1$ & $0.6 \leq \dfc < 3$ & $\bv < 1.4$ & $\{0.64,0.92,0.55\}$ & $\{0.83,0.96,0.61\}$ & $\{1.02,1.01,0.68\}$ \\
$< 1$ & $< 0.6$ & $< 0.6$ & 2.1 & 4.4 & 6.8 \\
$< 1$ & $< 0.6$ & $0.6 < \bv < 1.4$ & 2.3 & 3.5 & 5.3 \\
$< 1$ & $< 0.6$ & $> 1.4$ & 2.5 & 3.9 & 6.7\\
\enddata
\end{deluxetable}

\subsection{Other Regions\label{exceptions}}

Stars that are both active {\it and} either evolved or
very red ($\bv > 1.4$) are not discussed here, because there are too few
sample stars to properly determine the behavior of such stars.

Only two sample stars meet the criteria for study here
(\S~\ref{sample}) and have $\dfc > 0.6$ and $\bv > 1.4$.   They have
\sj\ values of 20 and 11 m\persec, both higher than the values predicted
by the median curve in Fig.~\ref{actjit}, which are $\sim 5$ m\persec.
This indicates that red, active stars may have substantially more jitter
than their bluer or more-inactive counterparts.

The region of \dfc-\dms\ space corresponding to active ($\dfc > 0.6$),
evolved ($\dms > 1$) stars is poorly populated, not just in this sample,
but also in other surveys of Sun like stars \citep{Wright2004b}.
The sole such star in this sample has $\sj = 21$ m\persec, substantially
higher than 
the 6 m\persec predicted by the \sj-activity relation here.  Such stars
often represent massive early F stars that have entered or 
are entering the subgiant phase before losing substantial amounts of
angular momentum \citep{do Nascimento2003}.

\section{The Jitter Model}

Table~\ref{jitters} contains the \sj\ values that are typical for
various regions given in (\bv)-\fc-\dms\ space.  The jitter model assumes that
a given star will exhibit levels of jitter similar to those of other
stars in its region.  When the Planet Search pipeline fits Keplerian
orbits to radial velocity time series, it incorporates the
jitter-model estimate of the jitter into its estimate of the errors.   
Furthermore, the model also consists of interpreting the median values in the
appropriate region of Table~\ref{jitters} as the estimated jitter, but
with the following minor modifications.

To prevent large discontinuities in the jitter model, the model
linearly interpolates the median jitter for stars within 0.1 mag of the
\bv\ boundaries.  For instance, an unevolved, inactive star with
$\bv=0.65$ would  have predicted jitter of 3.725 m\persec : since it is
within 0.1 mag of the boundary between the fourth and fifth entries in
Table~\ref{jitters}, the model interpolates between the median value for
the $\bv < 0.6$ and $0.6 < \bv < 1.4$ regions.  The model also
similarly interpolates between the first and second regions of
Table~\ref{jitters} for stars with $1.65 < \dms < 1.85$.  The
model also sets the predicted jitter of active stars ($\fc > 0.6$)
to the greater of the predicted jitter and that predicted for an
inactive star with the same \bv\ and \mv.

Both this jitter model and the regions and values in
Table~\ref{jitters} have evolved and become more sophisticated over
time, so the model described here may not agree precisely with jitter estimates
published in previous work by the California and Carnegie Planet Search.

This model is generally applicable for most Planet Search stars.
Since it was derived using Keck data, in particular data before the
September 2004 upgrade of the HIRES CCD, it accounts for systematic
errors that may not be present in radial velocity data from other
telescopes.  For jitter values $> 3 $m\persec, however, it
predominantly measures astrophysical jitter, and so is generally
applicable.  The presence of undetected planets around the sample
stars means that these values may be slightly overestimated, but the use of
medians mitigates this effect. 

Jitter is more important to the detection of exoplanets than the
characterization of their orbits.  The effects of including jitter in
one's noise estimate on best-fit orbital parameters for a
single-planet system are generally very small ($<1\%$), in part
because the 
errors are typically already nearly equal.  Jitter plays a more important role in
determining whether radial velocity variations are of stellar
or Keplerian origin.  To the extent that the jitter measured here is
astrophysical and not due to systematics or still-undetected planets,
the $2-5$ m \persec of jitter exhibited by most inactive stars
represents a minimum detection threshold below which planets will be
very difficult to detect.    

\section{Conclusions}

This work concerns a sample of the California and Carnegie Planet
Search data
stars representing stars without obvious or detected planets observed
at the Keck Observatory.  This sample describes the typical radial
velocity stability, measured as \sj, of most stars in the Planet
Search.  The metric \sj\ includes the effects of undetected planets,
systematic errors ($\lesssim 3$m\persec), and astrophysical jitter.

Among inactive stars F, G and K, and M dwarfs show sightly different
amounts of radial velocity jitter, from 3.5 m\persec for  G and K
dwarfs to nearly 5 m\persec for M dwarfs.  Evolved stars ($\dms  > 1.75$) show 
significant amounts of jitter, while those only slightly evolved
appear more consistent with main sequence stars.   

Active stars show markedly more jitter ($5-10$ m \persec) than
inactive stars, and this jitter increases with activity.  The
activity metric \fc\ of Rutten can provide an estimate of jitter
without the strong color dependence of previous models that employed \rphk\
as a metric.  

\appendix

\section{Medians and Percentiles of Data Sets That Include Limits\label{median}}
The median is defined as the $N/2$th largest element of a sorted 
$N$-element data set.  The presence of measurements for which one can
only set limits (nondetections, for instance) complicates matters
somewhat, but a median can still be similarly defined.

Consider a sorted $N$-element data set $x$, comprising elements
$x_i$ where $0 \leq i < N$.  Each $x_i$ represents a measurement that 
is either exact (a {\it data point}), an {\it upper limit}, or a {\it
  lower limit}.  The set $x$ should be sorted such that for otherwise equal
elements, an upper limit is less than a data point, which
in turn is less than a lower limit.  (Note that these definitions
apply only to the sorting procedure;  different comparison conditions
apply below.)  One seeks to test the value of each element $x_i$ as a
candidate median by comparing $a_i$, the number of elements unambiguously above
the value of $x_i$, to $b_i$, the number unambiguously below it.

For the purposes of computing $b_i$, element $x_j$ is counted as being
below the value of $x_i$ if, in 
the sorted array: (1) $x_j$ is a data point and $j < i$ or (2) $x_j$ is
an upper limit and $j \leq i$.  Note that when $x_i$ represents an
upper limit, $x_i$ is less than the value of $x_i$ by definition, and
so contributes to $b_i$, and that when $x_j$ represents a lower limit,
it does not contribute to $b_i$ at all.
Similarly, $x_j$ contributes to $a_i$ if $j > i$ when $x_j$ is a
data point, or if $j \geq i$ when $x_j$ is a lower limit.

The median is then is the value of the $i$th element of $x$ such that
$a_i$ is most 
nearly equal to $b_i$. If more than one such element exists, these can simply
be averaged.  In some cases, such as a data set consisting entirely
of upper limits, no median can be defined.  Similarly, the $p$th
percentile can be defined as the value of $x_i$ such that 
$b_i/(a_i+b_i)$ is most nearly $p/100$.  Naturally, the 50th percentile is
equivalent to the median.

For example, consider the appropriately sorted data set $x =
\{<1,5,5,<7,10,50\}$, in which the first and fourth elements represent
upper limits.  The values of $a$ for this set are $\{4,3,2,2,1,0\}$,
and similarly $b = \{1,1,2,4,4,5\}$.  Thus, the median of this set is
5 ($=x_2$, since $a_2 = b_2$), the 20th percentile is 1 ($=x_0$,
since $b_0/(a_0+b_0) = 1/5 = p/100$ when $p=20$), and the 80th
percentile is 10 ($= x_4$, since $b_4/(a_4+b_4) = 4/5 = p/100$ when
$p=80$). 


\acknowledgments
I am indebted to Geoff Marcy and the California and Carnegie Planet
Search for the data, guidance, and advice that made this work
possible.  I also thank the anonymous referee for the extremely timely
and constructive comments that improved this paper.

I wish to recognize and acknowledge the
very significant cultural role and reverence that the summit of Mauna
Kea has always had within the indigenous Hawaiian community.  We are
most fortunate to have the opportunity to conduct observations from
this mountain.

This research has made use of the SIMBAD database, operated at CDS,
Strasbourg, France, and of NASA's Astrophysics Data System
Bibliographic Services.  It was made possible by the generous support
of Sun Microsystems, the NSF, and NASA through grant NAG5-12090.

\end{document}